\documentclass[conference]{IEEEtran}
\def\BibTeX{{\rm B\kern-.05em{\sc i\kern-.025em b}\kern-.08em
    T\kern-.1667em\lower.7ex\hbox{E}\kern-.125emX}}
\usepackage{cite}
\usepackage{amsmath}

\usepackage{amssymb}
\usepackage{mathtools}
\usepackage{amsthm}
\usepackage{makecell}
\usepackage{colortbl}
\usepackage{subfig}
\usepackage{caption}
\usepackage{tabularx}
\usepackage{enumitem}
\usepackage[utf8]{inputenc} %
\usepackage[T1]{fontenc}    %
\usepackage{url}            %
\usepackage{hyperref}
\usepackage{booktabs}       %
\usepackage{amsfonts}       %
\usepackage{nicefrac}       %
\usepackage{microtype}      %
\usepackage[most]{tcolorbox}
\usepackage{xcolor}         %
\usepackage{lipsum}
\usepackage{graphicx}
\usepackage{subcaption}    
\usepackage{hyperref}      
\usepackage{cleveref}
\usepackage{array}
\usepackage{float}

\theoremstyle{plain}

\theoremstyle{definition}

\theoremstyle{remark}

\usepackage[textsize=tiny]{todonotes}
 
\makeatletter
\DeclareRobustCommand\onedot{\futurelet\@let@token\@onedot}
\def\@onedot{\ifx\@let@token.\else.\null\fi\xspace}

\makeatother

\newcommand{\ignore}[1]{}   %

\setlist[itemize]{leftmargin=1em}
\tcbset{
  myobservation/.style={
    colback=gray!8,      
    colframe=gray!60,    
    fonttitle=\bfseries\small,
    coltitle=black,
    boxrule=0.5pt,
    arc=1mm,
    left=2mm, right=2mm, top=1mm, bottom=1mm,
    before skip=8pt, after skip=8pt,
  }
}

\newtcolorbox{obsinsight}[1][]{myobservation, title=OBSERVATION --> INSIGHT, #1}

\newtcolorbox{observationbox}[1][]{
  colback  = gray!8,        %
  colframe = black!45,      %
  breakable,
  boxrule  = 1pt,
  left=0pt, right=0pt, top=0pt, bottom=0pt,
  sharp corners,
  enhanced,
before skip = 0.5em,         %
  after  skip = 0.5em,         %
  #1                       %
}

\newtcolorbox{insightbox}[1][]{
  colback  = gray!15,       %
  colframe = black!65,      %
  breakable,
  boxrule  = 1pt,
  left=0pt, right=0pt, top=0pt, bottom=0pt,
  sharp corners,
  enhanced,
before skip = 0.5em,         %
  after  skip = 0.5em,                %
#1
}

\begin{document}

\title{Position: Human Factors Reshape Adversarial Analysis in Human-AI Decision-Making Systems}

\author{
\IEEEauthorblockN{Shutong Fan}
\IEEEauthorblockA{\textit{Clemson University} \\
Clemson, USA \\
shutonf@clemson.edu}
\and
\IEEEauthorblockN{Lan Zhang}
\IEEEauthorblockA{\textit{Clemson University} \\
Clemson, USA \\
lan7@clemson.edu}
\and
\IEEEauthorblockN{Xiaoyong Yuan}
\IEEEauthorblockA{\textit{Clemson University} \\
Clemson, USA \\
xiaoyon@clemson.edu}
}

\maketitle

\begin{abstract} 
As Artificial Intelligence (AI) increasingly supports human decision-making, its vulnerability to adversarial attacks grows. However, the existing adversarial analysis predominantly focuses on fully autonomous AI systems, where decisions are executed without human intervention. This narrow focus overlooks the complexities of human-AI collaboration, where humans interpret, adjust, and act upon AI-generated decisions. Trust, expectations, and cognitive behaviors influence how humans interact with AI, creating dynamic feedback loops that adversaries can exploit. To strengthen the robustness of AI-assisted decision-making, adversarial analysis must account for the interplay between human factors and attack strategies.

\textbf{This position paper argues that human factors fundamentally reshape adversarial analysis and must be incorporated into evaluating robustness in human-AI decision-making systems.} To fully explore human factors in adversarial analysis, we begin by investigating the role of human factors in human-AI collaboration through a comprehensive review. We then introduce a novel robustness analysis framework that (1) examines how human factors affect collaborative decision-making performance, (2) revisits and interprets existing adversarial attack strategies in the context of human-AI interaction, and (3) introduces a new timing-based adversarial attack as a case study, illustrating vulnerabilities emerging from sequential human actions. The experimental results reveal that attack timing uniquely impacts decision outcomes in human-AI collaboration. We hope this analysis inspires future research on adversarial robustness in human-AI systems, fostering interdisciplinary approaches that integrate AI security, human cognition, and decision-making dynamics.
\end{abstract}

\begin{IEEEkeywords}
Human-AI Collaboration, Adversarial Analysis, Machine Learning Security, AI-assisted Decision-making
\end{IEEEkeywords}

\section{Introduction}

With rapid advancements in artificial intelligence (AI), an increasing number of tasks across diverse domains are performed with unprecedented efficiency and accuracy, benefiting human decision-making. For example, AI assists drivers in navigating complex road conditions with advanced safety features~\cite{garikapati2024autonomous,bagwe2022robustonrampmergingaugmented}, helps professionals refine documents through tools like ChatGPT~\cite{peng2023makingchatgptmachinetranslation,azaria2023chatgptremarkabletool}, and streamlines the recruitment workflow in human resources~\cite{isha2020}.
Despite these advancements, %
AI remains highly vulnerable to adversarial attacks — subtle input perturbations designed to mislead AI models while remaining imperceptible to humans~\cite{kurakin2017adversarialexamplesphysicalworld,8611298,fu2024vulnerabilities}. %
Such attacks pose significant challenges to AI-enabled decision-making, particularly in safety-critical applications.

While extensive research has examined adversarial attacks on fully automated AI systems, it often overlooks a crucial factor: {human involvement in AI-assisted decision-making systems}. 
In real-world applications, humans provide oversight, make critical judgments, and address broader societal considerations in AI deployment~\cite{pazzanese2020ethical, rehman2023exploring, ferrel2024}. %
We position that \textbf{human involvement fundamentally reshapes adversarial attack strategies and redefines robustness analysis in human-AI decision-making systems}, where is significantly underexplored. For instance, extensive research has investigated how to mislead autonomous driving systems, creating serious safety hazards~\cite{ibrahum2025deep,physgan2020}; however, human intervention can alter outcomes - humans may recognize anomalies of driving decisions and take corrective action, mitigating potential accidents~\cite{rangesh2021autonomous,grewal2024predicting}. Unfortunately, little attention has been given to how adversarial attacks can strategically exploit human cognitive and behavioral tendencies to influence or distort decision-making processes. Addressing these gaps is critical for developing AI systems that are not only resilient to adversarial threats but also aligned with human intuition and reasoning.

To support our position and provide concrete evidence, we conduct a systematic analysis of the existing adversarial attack and defense literature (2021–2025) to assess the extent of research addressing human-related concerns. We begin with the curated list of adversarial example papers maintained by Carlini\footnote{\url{https://nicholas.carlini.com/writing/2019/all-adversarial-example-papers.html}}, which includes $9,275$ articles. To narrow the scope, we apply a keyword-based filter using $15$ human-centric terms, identifying $5,719$ papers whose titles or abstracts contain at least one keyword. We then compute the semantic cosine similarity between each abstract and predefined textual descriptions of six widely recognized human-related categories. As shown in \Cref{table:human}, \textbf{only 9 of 9,275 papers (approximately 0.09\%) engage meaningfully with human-related topics} upon manual verification, underscoring how underexplored human factors remain in adversarial machine learning, despite their growing importance in human-AI systems. The detailed review process is presented in~\Cref{app:human-related_literature}.

\begin{table*}[!tb]
    \centering
    \caption{Human-Centric Topics within Adversarial AI Literature (2021–2025).}
    \label{table:human}
    \resizebox{0.8\linewidth}{!}{%
    \begin{tabular}{@{}lccccccc@{}}
        \toprule
        \textbf{Year} & \textbf{Total Paper} & \makecell{\textbf{Human} \\ \textbf{Perception}} & \makecell{\textbf{Trust in} \\ \textbf{AI System}} & \makecell{\textbf{Impact on} \\ \textbf{Decision-Making}} & \makecell{\textbf{Usability \&} \\ \textbf{User Experience}} & \makecell{\textbf{Human-} \\ \textbf{in-the-loop}} & \makecell{\textbf{Ethical \&} \\ \textbf{Social Implication}} \\
        \midrule
        2025 & 646  & 0 & 0 & 0 & 0 & 0 & 0 \\
        2024 & 2717 & 1 & 1 & 0 & 0 & 0 & 1 \\
        2023 & 2221 & 0 & 2 & 0 & 0 & 0 & 1 \\
        2022 & 1978 & 1 & 1 & 0 & 0 & 0 & 1 \\
        2021 & 1713 & 0 & 0 & 0 & 0 & 0 & 0 \\
        \bottomrule
    \end{tabular}
    }
\end{table*}

Within the limited research that incorporates humans into adversarial analysis, most studies implicitly treat humans and AI models as independent entities, whose responses to adversarial inputs are evaluated in isolation~\cite{elsayed2018adversarialexamplesfoolcomputer,qin2019imperceptible}. 
For instance, adversarial examples have been crafted to deceive both AI systems and time-constrained humans~\cite{elsayed2018adversarialexamplesfoolcomputer, veerabadran2023subtle, schneider2022concept}. These approaches primarily evaluate whether adversarial inputs can fool both human and AI systems in parallel, but critically, they assume that human decisions are made independently of prior AI predictions and that human feedback does not influence subsequent AI behavior. However, they overlook a critical aspect of real-world deployments: humans and AI systems often engage in collaborative decision-making, where humans rely on, interpret, and adapt to AI outputs over time. This interaction creates a tightly coupled system in which human behavior is influenced by prior AI performance, and vice versa. Ignoring this interdependence limits the understanding of how adversarial attacks propagate through human-AI teams and how trust dynamics evolve under adversarial pressure.

To bridge this gap, \textbf{we advocate for a comprehensive exploration of human factors in adversarial analysis, aiming to enhance the robustness of human-AI interactive decision-making systems.} Rather than acting as isolated variables, human factors, such as human self-confidence, task risk and task complexity, dynamically shape human reliance behaviors, which directly influence how and when users choose to rely on AI outputs and in turn determine the effectiveness and consequences of adversarial attacks. Understanding these interactions is essential for identifying system vulnerabilities and designing adaptive defenses. This position paper makes the following contributions:

\begin{itemize}
    \item \textbf{Human-AI Collaboration under Adversarial Attacks~(\Cref{sec:human-AI}):} We revisit human-AI collaborative decision-making in adversarial settings and conduct a comprehensive literature review to identify key human factors, such as trust, task risk, and time sensitivity, that affect AI model reliance in collaborative decision-making. These factors provide the foundation for understanding how adversarial attacks exploit human behavior in interactive decision pipelines.
    \item \textbf{Human-AI Decision-Making Robustness Framework~(\Cref{sec:pipeline}):} We introduce a novel framework that jointly models human trust and AI performance under adversarial conditions. It features two interconnected components: (1) {AI reliance assessment}, capturing how human trust evolves dynamically, and (2) {AI performance assessment}, reflecting human evaluation of AI outputs.
    \item \textbf{Timing-Based Adversarial Attacks~(\Cref{sec:timing-based}):} We propose a new class of adversarial attacks that exploit temporal dynamics in human-AI collaboration. In particular, we focus on the \textit{sequential decision-making tasks}, introducing a timing-based attack that maximizes impact by adapting to evolving human trust. To the best of our knowledge, this is the first work to explicitly leverage dynamic human trust as an attack surface in sequential human-AI decision-making.
    \item \textbf{Simulation-Backed Vulnerability Insights~(\Cref{sec:simulation}):} We simulate sequential decision-making tasks where attacks are strategically inserted at varying time points, measuring how different attack strategies affect human trust dynamics and system robustness. We demonstrate through simulation that attack effectiveness in human-AI systems exhibits non-monotonic behavior: more frequent attacks are not necessarily more effective. Instead, optimal timing, aligned with human reliance patterns, enables stronger influence with fewer perturbations.
\end{itemize}

\section{Revisiting Human-AI Collaborative Systems Through the Lens of Adversarial Robustness}
\label{sec:human-AI}
Adversarial examples have been recognized as a fundamental vulnerability in modern AI systems~\cite{8611298,goodfellow2014explaining}. Most prior work in this area focuses on perturbing inputs to fool a model in fully autonomous settings~\cite{chahe2023dynamic,schwinn2023adversarial}. These attacks typically assume a static decision-maker and evaluate success solely in terms of prediction changes. However, in many real-world applications (e.g., autonomous driving, medical diagnosis), humans remain in the loop as ultimate decision-makers. In such settings, adversarial impact is not limited to model misprediction; it also depends on how human users perceive, evaluate, and respond to AI outputs.

To understand adversarial vulnerabilities in these human-AI collaborative systems, we must go beyond model performance alone and consider how human cognitive, psychological, and contextual factors reshape the attack surface. The four key human factors commonly studied in human-AI interaction literature are shown in Table~\ref{tab:human_factors}. These include cognitive traits such as self-confidence~\cite{CHONG2022107018,ma2024arereallysureunderstanding}, task-related dimensions like risk~\cite{FAHNENSTICH2024108107,klein2023}, complexity~\cite{Dietvorst2014,Xu2020}, and time sensitivity~\cite{cao2023,Swaroop_2024}. These factors jointly shape when and how users choose to rely on AI recommendations.

Building on this foundation, we present four key observations about human-AI decision-making under adversarial conditions, each grounded in prior insights from human-AI interaction research:

\subsection{Observation 1: Model performance is the primary anchor of human trust.}
\label{O1}

Humans gain more trust in an AI system as its predictions become more accurate~\cite{Yin2019,Lai2019,Zhang_2020,bansal2021accurateaibestteammate,steyvers2022bayesian,Vaccaro_2024} (Table~\ref{tab:human_factors}). Even small, adversary-induced drops in observed performance (e.g., accuracy) will quickly erode that trust: after observing just a few conspicuous errors, users often hesitate to use the AI system on subsequent tasks~\cite{Lu2023,zhou2021machine,chong2023evolution}.

Because trust is performance-driven, the adversary attack does more than create mispredictions; it also signals unreliability, prompting users to disengage from future AI assistance. The stronger or more frequent the perturbations, the faster confidence collapses; once users stop relying on the model, the attack can no longer influence decisions, resulting in a self-limiting trust–performance feedback loop.

\begin{insightbox}
\textbf{Insight 1: Adversarial attacks trigger a self-limiting trust–performance feedback loop.}
\end{insightbox}

\subsection{Observation 2: Human reliance on AI is multifaceted.}
\label{O2}

\begin{table*}[!tb]
    \centering
    \caption{Recent Studies of Key Factors Affecting Human Reliance on AI Models Across Observations in~\Cref{sec:human-AI}.}
    \label{tab:human_factors}
    \resizebox{0.7\linewidth}{!}{%
        \begin{tabular}{lllc}

        \toprule
        \textbf{Factors} & \textbf{Description} & \textbf{Relevant Studies} & \textbf{Obs.} \\
        \midrule
        Model Performance & Accuracy of the AI model 
            & \cite{Yin2019,bansal2021accurateaibestteammate,steyvers2022bayesian,Vaccaro_2024} & 1 \\
            \midrule
        Self-Confidence & User's belief in own judgment 
            & \cite{Ma2023,CHONG2022107018,ma2024arereallysureunderstanding,Salimzadeh2023,chong2023evolution} & 2 \\
        Task Risk & Safety-critical context 
            & \cite{FAHNENSTICH2024108107,klein2023,camilli2021towards,Kwon_2020,Fuchs_2024} & 2 \\
        Task Complexity & Difficulty level of the task 
            & \cite{app132412989,Xu2020,Dietvorst2014,Salimzadeh2023,leskovar2022increasedcomplexityhumanrobotcollaborative} & 2 \\
        Task Time-Sensitivity & Urgency, decision speed 
            & \cite{cao2023,Swaroop_2024,shaikh2019alexaknowanythingimpact,rinott_2024,elsayed2018adversarialexamplesfoolcomputer} & 2 \\
            \midrule
        Past Experience & Historical cooperation satisfaction 
            & \cite{Lu2023,virvou2024virtsi,li2023modeling} & 4 \\
        \bottomrule
    \end{tabular}
    }
\end{table*}

A growing body of human-AI interaction research shows that human reliance on AI is multiplex~\cite{wang2020human,xu2023transitioninghumaninteractionai,vodrahalli2022uncalibratedmodelsimprovehumanai}. Model performance alone cannot explain when people accept or reject AI. Prior work has identified additional cognitive and situational factors (Table~\ref{tab:human_factors}): 
\begin{itemize}
    \item Cognitive factors
    \begin{itemize}
        \item Self-confidence. Self-confidence refers to the user's belief in their own judgment. This trait often varies by domain; individuals tend to exhibit higher self-confidence in areas where they possess expertise and lower confidence in unfamiliar domains~\cite{pajares1996self,dikmen2022effects}. Researchers state that individuals with high self-confidence are less inclined to defer to AI, even when the system is highly certain~\cite{CHONG2022107018,Lu2023,ma2024arereallysureunderstanding}. 
    \end{itemize}
    \item Situational factors
    \begin{itemize}
        \item Task risk. Task risk is defined as the safety-critical or context of the task. For example, in autonomous driving scenarios, driving through heavy rain at night carries a significantly higher risk than driving on a clear, quiet road. Prior works suggest that the same decision-makers become more willing to use AI when the cost of error is high~\cite{FAHNENSTICH2024108107,klein2023, camilli2021towards}. 
        \item Task complexity. Task complexity is the inherent difficulty level of the task. The impact of task complexity has been extensively studied. For instance, solving calculus problems typically involves higher complexity than performing basic arithmetic operations like addition or subtraction. However, prior findings on the relationship between task complexity and user trust remain mixed: some studies report greater trust for simple tasks~\cite{Xu2020, buccinca2021}, others for complex ones~\cite{Dietvorst2014}, and several document a U-shaped pattern with the lowest trust at moderate complexity~\cite{app132412989}. Recent work suggests that different dimensions of complexity (components, coordination, dynamics) may underlie these discrepancies~\cite{Salimzadeh2023}.
        \item Time sensitivity. Time sensitivity refers to the urgency of a task or the speed required for decision-making. For example, autonomous driving demands rapid responses, whereas using a large language model for literary translation allows more deliberation. Research shows that longer decision windows give users room to verify suggestions, increasing both acceptance and accuracy~\cite{cao2023,rastogi2022}. Under tight deadlines, people tend to follow the AI to make immediate decisions~\cite{Swaroop_2024,haduong2024raising}.
    \end{itemize}
\end{itemize}

These factors modulate human reliance, thereby reshaping the landscape for adversarial analysis. Therefore, adversarial attacks should consider more than just optimizing perturbations against AI models. The cognitive and situational factors listed above also play critical roles in determining attack effectiveness and strategies. Several illustrative examples follow.
\begin{itemize}
    \item Low-risk tasks (e.g., e-mail triage, movie picks). Users often accept model suggestions without scrutiny, lowering the chance that subtle manipulations or minor errors will be noticed. In such cases, adversarial perturbations can be more significant, and attackers can increase the frequency of attacks without detection.
    \item High-risk, slow-paced tasks (e.g., financial approval, non-urgent clinical reviews). The users inspect model output more closely when mistakes are costly~\cite{hunter2024monitoring,rosbach2025two}. An attacker must craft subtler perturbations. Once users detect mistakes, they may lose trust and abandon the AI model entirely, limiting the impact of the adversarial attacks.
    \item Time-critical tasks (e.g., autonomous driving, ER triage, real-time trading). Under time constraints, even in high-risk scenarios, users may bypass additional verification steps and rely entirely on the AI’s output~\cite{zhai2024effects}. This creates a window of vulnerability that attackers can exploit.
\end{itemize}

These scenarios show that human factors other than model performance can dampen or magnify adversarial effects, create new attack paths (e.g., exploiting workload spikes), and reshape the adversarial landscape. Adversarial analysis must therefore consider both cognitive and situational factors.

\begin{insightbox}
\textbf{Insight 2: Cognitive and situational factors shift where, when, and how an attacker can succeed.}
\end{insightbox}

\subsection{Observation 3: Humans take actions in the decision making; the reliance shapes human actions.}

\label{O3}
Humans actively participate in the decision-making pipeline; they are not passive recipients of AI output. Once trust is established, it drives concrete human behavior.  In the decision-making loop, humans continually determine whether to watch the AI, when to step in, and how to correct it~\cite{hunter2024monitoring}. 
The amount of trust humans have in the AI system triggers different actions. For instance, clinicians may choose to pay more attention to the AI suggestions and tend to reject the suggestions when they detect frequent inconsistencies with patient context or prior knowledge~\cite{sivaraman2023, buccinca2021,fogliato2022,zhou2021machine}. 

These meta-decisions introduce an additional layer of defense, creating robustness that adversarial attacks must circumvent. An attack succeeds only if the incorrect predictions still result in the final decision. Any adversarial analysis that ignores these human safeguard actions will overstate how much damage an attacker can actually cause.

\begin{insightbox}
\textbf{Insight 3: Adversarial attacks that fool the AI model are only halfway successful.}
\end{insightbox}

\subsection{Observation 4: Human-AI decision-making systems are interactive and unfold over time.}

\label{O4}
Unlike conventional AI-only systems, where each decision is isolated and independent, human-AI collaboration forms a sequential process, in which past interactions shape future decisions~\cite{Lu2023, li2023modeling,virvou2024virtsi} (Table~\ref{tab:human_factors}). 
Users update their reliance after each success or failure; then, reliance determines their actions, e.g., how closely they will monitor, accept, or override the model in the following cases. This feedback loop turns even ``single-step'' tasks (e.g., bail decisions, medical diagnoses) into a sequential process: a judge who consults the AI across several hearings, or a clinician who revisits an AI assistant across patients, accumulates experience that shifts future reliance on AI. The result is a moving target for both accuracy and vulnerability, different from an isolated AI system.

The temporal structure of human-AI decision making introduces new attack surfaces: adversaries can not only target individual predictions, but also manipulate the dynamics of human trust and reliance as they evolve. If the model is accurate for the first few tasks, users relax their checks. A significant error can then pass unnoticed, transforming a minor perturbation into a high-impact failure. Consequently, the objectives for both attackers and defenders should shift from success in isolated tasks to cumulative effectiveness over entire interaction sequences.

\begin{insightbox}
\textbf{Insight 4: Adversarial analysis must consider the full sequence of interactions, rather than a single isolated decision point.}
\end{insightbox}

\section{Robustness Analysis Framework for Human-AI Decision-Making} %
\label{sec:pipeline}

\subsection{Overview}

\begin{figure}[!tb]
\centering
\includegraphics[width=1\linewidth]{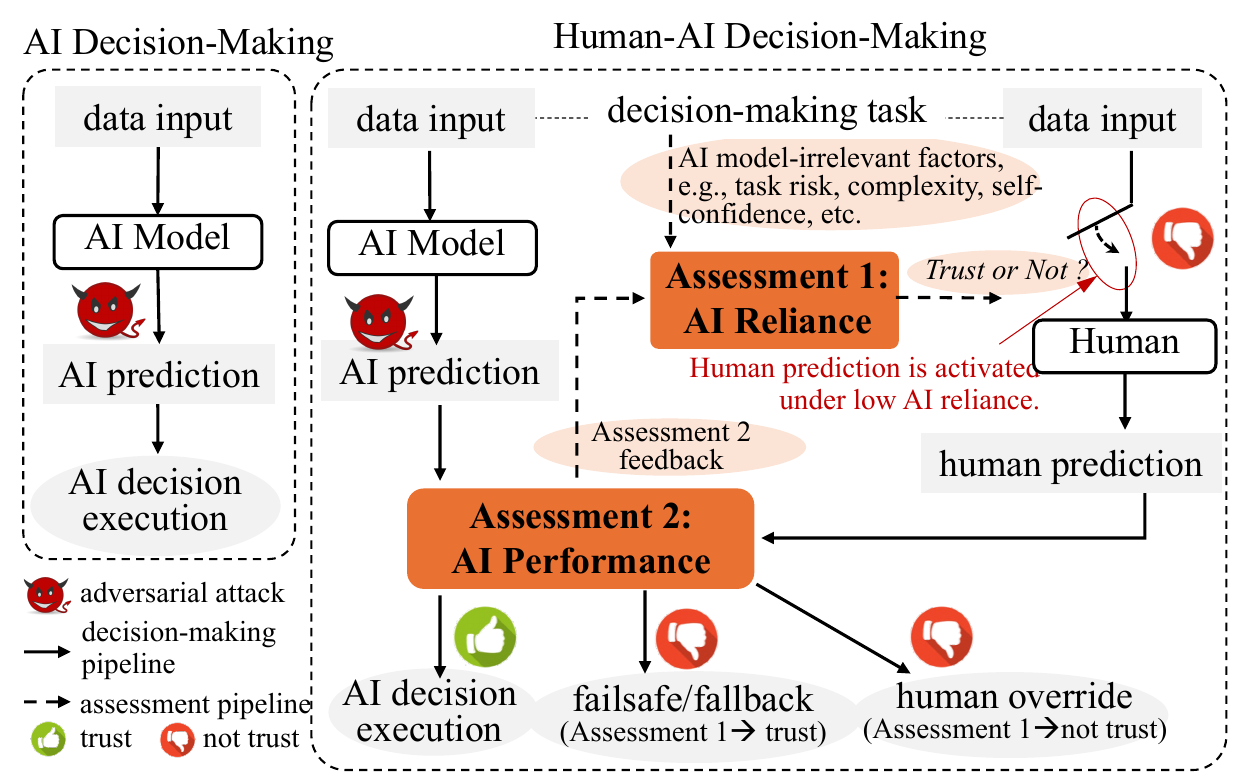}
\caption{Comparison between conventional AI-only decision-making (left) and our proposed robustness analysis framework for human-AI decision-making (right).}
\label{pipeline}
\vspace{-1em}
\end{figure}

To better understand the evolving role of humans in interactive decision-making with AI, we introduce a robustness analysis framework that models human behaviors under adversarial conditions. Inspired by the four observations derived from our analysis of model performance and human factors influencing reliance behavior, our framework is structured around two core assessments: \textbf{AI reliance assessment} and \textbf{AI performance assessment}, as illustrated in~\Cref{pipeline}. The \textit{AI reliance assessment} captures how users decide whether to trust AI predictions, based on both past AI performance and model-irrelevant factors, which are task-specific cognitive and situational factors (e.g., human self-confidence, task complexity etc.). The \textit{AI performance assessment} evaluates the accuracy of AI outputs through human feedback and, in turn, influences future trust and governs whether to execute, override, or fall back. These two assessments form a dynamic feedback loop that reflects the temporal, history-dependent nature of human-AI interaction. Together, they capture how humans interact with AI predictions, update their trust, and respond to adversarial threats within this temporally extended, sequential setting.

\subsection{Assessment 1: AI Reliance}
\label{reliance_assessment}

AI reliance assessment determines whether a human chooses to trust an AI model's prediction. As discussed in~\Cref{sec:human-AI}, this decision is shaped by two primary components: (1) AI performance feedback, reflecting prior experiences with the model, and (2) AI model-irrelevant factors, such as cognitive and contextual factors (e.g., task risk, self-confidence). Together, these two components jointly shape human reliance decisions.

\begin{itemize}
    \item \textit{AI performance feedback} – Derived from the AI performance assessment below, AI performance captures historical performance trends in interactive human-AI collaboration, which dynamically influences future reliance levels. Analogous to human–human collaboration, a poor prior experience lowers the likelihood of partnering ~\cite{bohnet2004trust,bansal2015}.~\cite{baumeister2001bad} points out that negative experiences weigh more heavily than positive ones in relationships and future choices. Human–AI collaboration follows the same pattern that dissatisfying outcomes reduce willingness to rely on the AI model, while consistent successes rebuild trust.
    
    \item \textit{AI model-irrelevant factors} – These include psychological and situational variables such as self-confidence, perceived risk, task complexity, and time sensitivity, which influence trust independently of the AI's actual performance (as discussed in \Cref{O4}). In addition to considering the AI model’s historical performance, humans also evaluate these model-irrelevant factors for each task to ultimately decide whether to adopt the AI’s predictions. For example, although some people believe in AI models with perfect historical performance, they often prefer their own judgment in extremely high-risk tasks (e.g., surgery, air-traffic control, safety-critical driving maneuvers), deferring less to the AI~\cite{bigman2018people,he2022modelling}.
\end{itemize}

When reliance conditions are met, the human defers to the model without making their own prediction. Otherwise, they generate an independent decision. Existing work has extensively examined trust in AI~\cite{kaplan2023trust, Benk_2024, maier2022relationship,cao2022understanding,schoeffer_2023,He_2023}, yet most studies overlook the role of past AI performance in shaping trust. Our framework addresses this gap by incorporating historical AI performance as a dynamic factor, allowing trust to evolve over time.

\subsection{Assessment 2: AI Performance }
\label{performance_assessment}

AI performance assessment evaluates the accuracy and reliability of AI predictions in two scenarios: when humans 1) trust or 2) do not trust AI.

\begin{itemize}
    \item \textit{When humans do not trust AI (low reliance in assessment 1)}: When humans do not defer to the AI, humans generate their own predictions and compare them to AI model's outputs, referred to as human-model prediction assessment. Evaluation methods vary from binary correctness checks (match/mismatch) to metric-based assessments (e.g., BLEU~\cite{papineni2002bleu}, ROUGE~\cite{lin-2004-rouge}, edit distance~\cite{przybocki-etal-2006-edit}). For example, a driver who uses the autonomous driving technique on the road has low reliance on the autonomous driving system; the human is supposed to verify the stop sign and intervene if the system starts to proceed through the intersection

    \item \textit{When humans trust AI (high reliance in assessment 1)}: In most cases, humans adopt the AI's predictions directly, resulting in the AI decision execution. However, even when relying on AI, humans may still assess predictions indirectly, known as model-only prediction assessment. For example, in autonomous driving, an attack could cause an AI model to misclassify a stop sign. A driver might not visually check the sign but would notice unexpected acceleration and intervene.

\end{itemize}

This assessment serves two primary functions. One is \textit{decision execution}. If the AI's prediction is deemed reliable, it is executed. 
Otherwise, human intervention overrides the AI decision when a human prediction is available. If no human prediction is available (i.e., the AI model is trusted), a failsafe or fallback mechanism is triggered. 
Another one is the \textit{feedback loop}. Performance evaluation results influence future AI reliance assessments, ensuring trust is updated dynamically based on AI reliability. Consistent with prior research~\cite{kaplan2023trust, Benk_2024,schoeffer_2023}, trust in AI increases with positive experiences and decreases with observed failures, leading to more intensive human assessment.

\subsection{Human-Aware Adversarial Attack Strategy}
\label{attack_strategy}

Building on the dual assessments of AI reliance and performance introduced in our framework, we now examine how adversarial strategies must adapt in the context of human-AI collaboration. Traditional adversarial attacks are designed for AI-only decision-making, where models execute predictions without human oversight. These attacks optimize for single-task disruptions, aiming to mislead AI outputs as efficiently as possible. However, in human-AI interactive systems, attackers must also account for the role of humans: how human monitoring and changing trust. Our framework introduces two innovative adversarial considerations:

\begin{itemize}
\item \textit{Human Awareness of Anomalies}: Unlike fully automated pipelines, human users may detect adversarial perturbations even without direct input observation. For example, in autonomous driving, a driver might not see a manipulated stop sign but could recognize anomalous acceleration and override the AI.

\item \textit{Adaptive Trust Exploitation}: Adversaries must strategically balance the impact of their attacks with the need to avoid rapid trust degradation to sustain long-term influence.
Instead of consistently triggering obvious errors erode human reliance on AI, attackers may selectively introduce perturbations, adapting to the user's evolving trust model to maintain influence over decision-making. Take an example of driving with the autonomous driving algorithm, rather than attacking indiscriminately, the adversary targets a subset of tasks conditioned on the user’s trust level, striking which tasks to attack to maximize expected threat effectiveness.

\end{itemize}

These factors force adversaries to go beyond conventional single-task attacks and consider the evolving trust dynamics between humans and AI.

\subsection{A Case Study on Autonomous Driving}
\label{case:autodriving}
To illustrate our framework, we analyze a self-driving scenario where \textit{an autonomous vehicle approaches a STOP sign}. 

\textbf{AI reliance assessment} determines whether the driver trusts AI predictions. If AI has historically misclassified traffic signs or demonstrated inconsistencies in object recognition, the driver may be more cautious and less likely to trust the AI's decision. Conversely, consistent and accurate past performance may increase trust. Additionally, even if AI has performed reliably, model-irrelevant factors influence trust. A driver's self-confidence plays a role - more experienced drivers may be less inclined to rely on AI, while those less confident in their driving skills may be more inclined to defer to the AI. Similarly, task risk level affects trust, such as drivers navigating in high-speed or high-traffic environments may be more hesitant to trust AI without verification. Task complexity also matters, as straightforward driving scenarios (e.g., an open highway) may lead to higher AI reliance, while complex environments (e.g., urban intersections with pedestrians and cyclists) may encourage greater human oversight.

If the driver trusts AI (i.e., high reliance), the model's interpretation of the stop sign is accepted, and no human decision is made. Otherwise, the driver may actively assess the stop sign against the AI prediction (i.e., low reliance). Once the AI and human predictions are available, they undergo \textbf{AI performance assessment} to verify the AI model's output. If the AI correctly identifies the stop sign and aligns with the driver's expectations, the vehicle safely comes to a stop. However, if an adversarial attack causes the AI to misinterpret the stop sign as a speed limit sign, a discrepancy emerges. The driver, if engaged in monitoring, may override the AI decision by manually applying the brakes. Alternatively, if AI performs without human intervention and misclassification goes undetected, the vehicle continues accelerating, potentially resulting in an accident.

This example illustrates how adversarial attacks can disrupt AI reliance and human-AI decision-making. Even when detected, these attacks can erode trust over time, leading drivers to either persistently override AI decisions, even when correct, or become overly reliant, failing to intervene when necessary. These findings highlight the need for adversarially resilient human-AI systems that account for both AI performance and human trust dynamics. By integrating historical AI performance, cognitive factors, and task context, our framework provides a structured approach to understanding how adversarial attacks manipulate human-AI decision-making.

\section{Timing-based Adversarial Attacks against Human-AI Decision-Making: A Case Study}
\label{sec:timing-based}
Human-AI collaborative decision-making system introduces a sequential process in which tasks are interdependent and prior successes inform subsequent performance as discussed in~\Cref{O4} and illustrated in our robustness analysis framework~\Cref{sec:pipeline}. This dynamic creates a new adversarial surface where the timing of perturbations can shape long-term system vulnerability. To explore this, we introduce a timing-based adversarial attack as a case study that exploits human behavioral dynamics across task sequences. Specifically, we investigate how adversaries can strategically target certain tasks to maximize cumulative system risk. For clarity, we summarize all variables involved in the analysis in~\Cref{tab:vars-timing}.

\subsection{Threat Model}
\label{sec:threat model}
\subsubsection{Adversary's Goal}

We consider a human-AI decision-making system consisting of $n$  sequential tasks $\{T_i\}, i=1,2, \cdots, n$. The adversary aims to determine the optimal set of tasks to perform attacks, represented as \(\mathbf{a} = (a_1, a_2, \dots, a_n), \quad a_i \in \{0, 1\}\), for maximizing the overall Attack Score~(\(\mathrm{AS}\)). We define \(\mathrm{AS}\) as a metric to quantify the attack success, where higher values indicate greater adversarial impact. 
The per-task attack score is given by 
\begin{equation}
    \mathrm{AS}_i = \mathcal{L}(y_i, y_i'),
\label{eq:as}
\end{equation}
where  $\mathcal{L}$ is a discrepancy function that measures the deviation between the ground truth  $y_i$  and the actual system prediction $y_i'$. This function can be instantiated using metrics such as absolute difference or cross-entropy loss, depending on the prediction task.

Attackers with different objectives may adopt distinct methods for computing the overall attack score \(\mathrm{AS}\). 
For instance, in binary classification settings with high fault tolerance, the system is only considered compromised if all tasks are misclassified: 
\begin{equation}
    \mathrm{AS} = \prod_{i} \mathcal{L}(y_i, y_i'),,
\label{eq:attacker_goal_prod}
\end{equation}
where the attacker must ensure that: $y_i \neq y_i'$ for all tasks to fully deceive the system. In the following experiments, we assume each task is equally important and compute the overall attack score as the average of per-task scores:

\begin{equation}
\label{eq:sum_as}
\mathrm{AS} = \frac{1}{n}\sum_i^n \mathrm{AS}_i.
\end{equation}

\subsubsection{Adversary's Capability} 

We consider a timing-based attack strategy, where the adversary chooses which tasks to perform attacks. The attack strategy is denoted as $a_i$. $a_i=1$ indicates the adversary performs an attack on task $\mathcal{T}_i$, $a_i=0$ otherwise. 
To explore the worst-case robustness of human-AI systems, we assume the white-box attack, where the attacker has full knowledge of the human-AI decision-making pipeline as well as both assessment ~\cite{goodfellow2014explaining,madry2019deeplearningmodelsresistant}.

However, we assume that adversarial attacks cannot deceive human predictions~\cite{elsayed2018adversarialexamplesfoolcomputer,veerabadran2023subtle}. That is, while the attack can manipulate AI outputs, human predictions are not affected by adversarial attacks. 

\subsubsection{Target Human-AI System} 

 Following the proposed pipeline in \Cref{pipeline}, we define the system settings as follows:

\noindent\textbf{AI reliance assessment:} At each task $T_i$, we define a reliance score $r_i$ that quantifies the human reliance in the AI model based on (1) AI performance feedback $D_i$ derived from the model performance from the previous task and (2) model-irrelevant factors $I_i$ (e.g., task risk, complexity, self-confidence, etc.):
    \begin{equation}
        r_{i} = \gamma D_i + (1-\gamma) I_{i},
    \label{eq:reliance_new}
    \end{equation}
    where $\gamma$ balances two factors. Given that human reliance evolves over time, we incorporate a momentum-based update mechanism~\cite{RumelhartHintonWilliams1986}: \begin{equation}
    r_{i+1}^* = \alpha r_i^* + (1 - \alpha) r_{i+1},
    \label{eq:reliance_update}
    \end{equation}
where \(\alpha\) controls the balance between historical reliance score \(r_{i}^*\) and the newly computed reliance score \(r_{i+1}\). Based on the updated reliance score  $r_{i+1}^*$, the human makes a trust decision: if $r_{i+1}^* \ge \hat{r}$, the AI model is trusted. Otherwise, the AI model is not trusted, and the human provides their own prediction.

\begin{itemize}
    \item \noindent Model-irrelevant factors: In model-irrelevant factors \(I_i\), we consider four key factors as described in~\Cref{O4}: 1) \textit{Human self-confidence $C_i$}: Higher self-confidence has been shown to decrease the likelihood of accepting AI model predictions~\cite{chong2023evolution,ma2024arereallysureunderstanding}. We model this inverse relationship as: 
    \begin{equation}
        C_i = - w_c c_i,
    \label{eq:self-conf}
    \end{equation} 
    where $w_c$ controls the influence of self-confidence $c_i$ on reliance. 2) \textit{Task risk $K_i$}: Higher-risk tasks tend to make humans more cautious when relying on AI, as observed in~\cite{lubars2019askaidoai}. To capture this effect, we use an exponential decay function to model the rapid decline in trust for high-risk tasks: 
    \begin{equation}
        K_i = w_k e^{-k_i},
    \label{eq:risk}
    \end{equation}
    where \( w_k \) determines the influence of task risk \( k_i \) on reliance. 3) \textit{Task complexity $O_i$}: Research suggests that as task complexity increases, reliance on AI initially grows but may decline if the complexity becomes overwhelming~\cite{Salimzadeh2023}, we approximate this behavior with a quadratic function:
    \begin{equation}
        O_i = w_o o_i^2,
        \label{eq:complexity}
    \end{equation}
    where \( w_o \) is a hyperparameter that quadratically influences the time-sensitivity \( o_i \) on reliance. 4) \textit{Time sensitivity $S_i$}: Under time constraints, humans tend to rely more on AI assistance to expedite decision-making~\cite{cao2023}. We model this direct relationship as: 
    \begin{equation}
        S_i = w_s s_i,
        \label{eq:sensitivity}
    \end{equation}
    where \( w_s \) is a hyperparameter and  \( s_i \)  denotes the time-sensitivity on reliance. The final model-irrelevant factor \( I_i \) is computed in our experiment as: 
    \begin{equation}
    I_i = C_i + K_i + O_i + S_i.
    \label{model-irrelavant}
    \end{equation}
    \item \noindent AI performance feedback:
We consider two measurements for AI performance: 1) model prediction measurement, when the AI model is trusted but no human prediction is available: \begin{equation}
D_i = c\, d_i^m.
\label{performance_assessment_model}
\end{equation}
\(d_i^m\) is the evaluation score in model prediction , where $c$ is the scaling coefficient. 2) human-model prediction measurement, where the human thinks the AI model is not trusted, and a human prediction is provided: 
\begin{equation}
D_i = c\, d_i^h.
\label{performance_assessment_human}
\end{equation} \(d_i^h\) is the evaluation score in human-model prediction measurement.
\end{itemize}

\noindent\textbf{AI performance assessment:} 
From the proposed framework, we can find that the attack score is directly determined by the actual system prediction $y_i',$
Humans are deemed satisfied when the human evaluation score without human prediction $d^m$ equals or exceeds the threshold for model-only prediction assessment $\theta^m$, represented as \( d^m \geq \theta^m \); otherwise, they are considered unsatisfied. Similarly, humans are considered satisfied when the human evaluation score with human prediction $d^h$ meets or exceeds the threshold for human-model prediction assessment $\theta^h$, expressed as \( d^h \geq \theta^h \). Otherwise, they are deemed unsatisfied.

\subsection{Human Trust-Gated Decision on Adversarial Attacks}

In AI-only decision-making systems, adversary can deterministically manipulate the model's prediction  $y'$, resulting in misleading outputs. However, in human-AI interactive systems, the prediction $y'$ depends on human assessment results, introducing variability into the decision-making process.

We define the predictions as follows. $y_i^M$ denotes the AI model's prediction without attacks, $y_i^A$ the prediction under attacks, $y_i^H$ human prediction, and $y_i^F$ failsafe/fallback prediction. Following our decision-making pipeline in \Cref{pipeline}, we consider four possible scenarios based on AI reliance and AI performance assessment results:
\begin{itemize}
\item AI model is trusted, and the model's prediction passes the performance assessment, humans directly adopt the model's prediction. $y_i'=y_i^M$ if no attack is launched ($a_i=0$), $y_i' =y_i^A$ otherwise.
\item AI model is trusted, but the model's prediction fails the performance assessment: humans revert to the failsafe or fallback mechanism: $y_i' = y_i^F$.
\item AI model is not trusted, and the model's prediction passes the performance assessment. Humans verify and accept the model prediction: $y_i'=y_i^M$ if no attack is launched ($a_i=0$), $y_i' =y_i^A$ otherwise.
\item AI model is not trusted, but the model's prediction fails the performance assessment: humans override the model prediction by human prediction: $y_i'=y_i^H$.
\end{itemize}

By updating $y_i'$ with four different scenarios, the adversary aims to achieve the best attack performance by maximizing the objective function~Eq.~\ref{eq:sum_as}.

\subsection{When to Strike: Timing-Based Attack Strategy Analysis}
\label{case study}

This section examines the effectiveness of timing-based adversarial attacks by comparing different attack strategies and showcases the unique characteristics of adversarial attacks in human-AI decision-making systems compared with AI-only systems.

To better analyze adversarial attack strategies, we simplify the target system with the following assumptions. 1) No failsafe or fallback mechanism is adopted. The fallback prediction defaults to the attacked model prediction $y_i^F=y_i^A$. 2) If the AI model is not trusted, performance assessment always fails, leading to the human prediction: $y_i' = y_i^H$.
Under these assumptions, the final prediction simplifies to:
\begin{equation}
y_i' =
\begin{cases} 
    a_i y_i^A + (1 - a_i) y_i^M & \text{if } r_i^* \geq \hat{r}, \\
    y_i^H & \text{if } r_i^* < \hat{r}.
\end{cases}    
\end{equation}
3) The initial reliance value \(r_1^*\) is high, ensuring that the AI model will always be trusted in the first task. 4) We choose a low reliance update weight \( \alpha \) in~Eq.~\ref{eq:reliance_update} to prioritize the current reliance score over its historical average, which better mirrors mostly real-world reliance dynamics.
5) Prediction errors remain consistent across all tasks. Then we simplify the errors by new notations: $e^H = \mathcal{L}(y_i, y_i^H), e^M = \mathcal{L}(y_i, y_i^M), e^A = \mathcal{L}(y_i, y_i^A)$. 6) To emphasize the dynamics of AI performance feedback, we isolate the effect of AI performance feedback without the model-irrelevant factors in this analysis: $I_i = 0$.

To examine how attack timing influences adversarial success, we consider two cases: one-time attack, where an adversary can only launch a single attack across $n$  tasks, i.e., $\sum_{i=1}^{n} a_i = 1$ and two-time attack, where an adversary can only attack twice across $n$ tasks, i.e., $\sum_{i=1}^{n} a_i = 2$.

\subsubsection{One-time Attack}
In the one-time attack, we compare the attack performance of two attack strategies.

\noindent\textbf{Attack Strategy 1:} The adversary launches an attack at the first task: $a_1 = 1, a_2 = 0, \dots, a_n = 0$.
\label{attack_str1}

Assuming the AI model is initially trusted at task \(T_1\) (\( r_1^* \geq \hat{r} \)), then the updated reliance value is given by: \(r_2^* = \alpha r_1^* + (1 - \alpha)\, c\, d_1,\) by~Eq.~\eqref{eq:reliance_new}-\eqref{eq:reliance_update}. Since an attack occurs at \(T_1\), the evaluation score of task 1 \( d_1 \) is expected to be small. With a small reliance update weight \( \alpha \), it yields that \( r_2^* < \hat{r}\), so the model is not trusted and the human prediction is used,  \(y_2' = y_2^H\). The reliance of task $n$ can be derived as: 
\begin{equation}
    r_n^* = \alpha^{n-1} r_1^* + c (1 - \alpha) \sum_{k=1}^{n-1} \alpha^{n-1-k} d_k.
\end{equation}
Accordingly, the same pattern propagates to \(r_3^*, \ldots, r_n^*\). Under this trajectory, the attack score becomes:
\(\mathrm{AS} = \frac{1}{n}\left(e^A + (n-1) e^H\right),\) where \(e^{A}\) and \(e^{H}\) denote the per-task loss under an effective attack and under human-only decisions, respectively.

\noindent\textbf{Attack Strategy 2:} The adversary waits until the final task to launch the attack: $a_1 = 0, \dots, a_{n-1} = 0, a_n = 1$.
\label{attack_str2}

Suppose the human initially trusts the AI model (\( r_1^* \geq \hat{r} \)), then \( y_1' = y_1^M \). With the update \(r_2^* = \alpha \cdot r_1^* + (1 - \alpha)\, c\, d_1,\) it follows that \( r_2^* \geq \hat{r},\) hence \(y_2' = y_2^M \). This pattern persists similarly for \( r_3^*, \dots, r_n^* \). The attack score becomes: \(\mathrm{AS} = \frac{1}{n} \left(e^A + (n-1) e^M\right) .\)

The key difference between the two strategies lies in the loss under human versus model: $e^H$ vs. $e^M$. The results suggest that if humans are generally more accurate than the AI model, i.e., \(e^H < e^M\), then delaying the attack until the final task leads to greater attack success.

\subsubsection{Two-time Attack}

We evaluate three attack strategies where the adversary launches two attacks over $n$ tasks.

\noindent\textbf{Attack Strategy 1:} The adversary launches attacks at the first two tasks: \(a_1=1,\, a_2=1,\, a_3=\cdots=a_n=0\).

Similar to \Cref{attack_str1}, after the first attack, assume initial trust (\(r_1^* \ge \hat{r}\)). The first attack makes the performance assessment for \(T_1\) small, then \((d_1\) is low. As a result, \(r_2^* = \alpha r_1^* + (1 - \alpha) c d_1\) will be reduced, leading to \( r_2^* < \hat{r} \). Humans override AI predictions: \( y_2' = y_2^H \). This pattern continues for all subsequent tasks. The final attack score becomes \(\mathrm{AS} = \frac{1}{n} \left(e^A +(n-1)e^H\right)\).

\noindent\textbf{Attack Strategy 2:} The adversary attacks the last two tasks: \(a_1=\cdots=a_{n-2}=0,\, a_{n-1}=1,\, a_n=1\).

Since attacks occur at the last two tasks, decisions follow the model up to \(T_{n-2}\). At \(T_{n-1}\) the model is trusted, so the attack yields \(y_{n-1}'=y_{n-1}^{A}\). That attack depresses the performance assessment for \(T_{n-1}\) $d_{n-1}$ is small and reliance update will be reduced: \(r_{n}^* = \alpha r_{n-1}^* + (1 - \alpha)\,c\, d_{n-1}\). As a result, humans override the AI decision in the last task \(y_n' = y_n^H\). Then the attack score \(\mathrm{AS} = \frac{1}{n}\left((n-2)e^M+ e^A + e^H\right).\)

\noindent\textbf{Attack Strategy 3:} The adversary attacks both the first and last tasks: \(a_1=1,\, a_2=\cdots=a_{n-1}=0,\, a_n=1\).

Suppose humans initially trust the AI model (\( r_1^* \geq \hat{r}\)). The first attack makes \(d_1\) small, and with small \(\alpha\), the reliance can be updated as
\(r_2^* = \alpha  r_1^* + (1 - \alpha)\,c\,d_1.\) So humans override the AI prediction immediately: \( y_2' = y_2^H \). Afterward, there are no attacks in the tasks in the middle part, so the performance assessment score among the middle task \( d_i \) would be larger. According to the reliance update rule \(r_{i+1}^* = \alpha r_i^* + (1 - \alpha) r_{i+1}\), there exists a task \(T_k, k\in\{2,\ldots,n-2\}\) in the middle where the reliance value after this task becomes greater than AI reliance assessment threshold \( r^* \), denoting the \emph{last} task at which the model is not trusted (\(y_k'=y_k^{H}\)); from \(T_{k+1}\) through \(T_{n-1}\) the model is trusted. The attack score becomes: \(\mathrm{AS} = \frac{1}{n}\left[e^A + (k-1)e^H + (n-k-1)e^M + e^A\right].\)

Comparing the two-time attacks at different attack timing in attack strategies 1, 2 and 3. The first two scenarios, which attack twice continuously at the beginning and twice at the end, are similar to those previously discussed in \Cref{attack_str1}, where the outcome primarily depends on the accuracy of the model prediction \(y_i^M\) and the human prediction \(y_i^H\). However, in attack strategy 3, where the attack occurs once at the beginning and once at the end separately, there is a possibility for the human to regain trust in the model during the middle tasks. This regained trust may lead the human to rely on the model’s outputs, making the final attack more likely to succeed \(y_n' = y_n^A\). Therefore, attacking once at the beginning and once at the end generally results in a stronger overall attack compared to the other two scenarios.

\begin{table*}[t]
\centering
\caption{Summary of variables introduced in the paper.}
\small
\setlength{\tabcolsep}{6pt}
\renewcommand{\arraystretch}{1.12}
\begin{tabularx}{\textwidth}{@{} l l X l @{}}
\toprule
\textbf{Symbol} & \textbf{Name} & \textbf{Description} & \textbf{Where} \\
\midrule
\multicolumn{4}{@{}l}{\textit{Tasks and attack plan}}\\
$n$ & \# tasks & Number of sequential tasks in the pipeline. & Sec.~\ref{sec:timing-based} \\
$T_i$ & Task $i$ & The $i$-th task in the sequence ($i=1,\dots,n$). & Sec.~\ref{sec:timing-based} \\
$a_i$ & Attack indicator & Whether task $T_i$ is attacked ($a_i\!\in\!\{0,1\}$). & Threat Model \\
\midrule
\multicolumn{4}{@{}l}{\textit{Predictions and losses}}\\
$y_i$ & Ground truth & True label/output for task $T_i$. & Eq.~\eqref{eq:as} \\
$y_i'$ & Executed prediction & System’s final decision on $T_i$ (after assessment). & Eq.~\eqref{eq:as} \\
$y_i^M$ & Model prediction (clean) & Model prediction without attack. & \S\;Human Trust-Gated Decision \\
$y_i^A$ & Model prediction (attacked) & Model prediction under attack. & \S\;Human Trust-Gated Decision \\
$y_i^H$ & Human prediction & Human prediction (when AI not trusted / override). & \S\;Human Trust-Gated Decision \\
$y_i^F$ & Failsafe/fallback prediction & Conservative fallback decision when assessment fails. & \S\;Human Trust-Gated Decision \\
$\mathcal{L}(\cdot,\cdot)$ & Discrepancy & Loss measuring deviation between $y_i$ and $y_i'$. & Eq.~\eqref{eq:as} \\
$\mathrm{AS}_i$ & Per-task attack score & Attack impact on task $T_i$. & Eq.~\eqref{eq:as} \\
$e^H,e^M,e^A$ & Error simplification & Per-task losses for human, clean model, attacked model. & \S\;\ref{case study} (assumptions) \\
\midrule
\multicolumn{4}{@{}l}{\textit{AI Reliance Assessment}}\\
$r_i$ & Instantaneous reliance & Reliance value for task $T_i$. & Eq.~\eqref{eq:reliance_new} \\
$\gamma$ & Feedback weight & Balances performance feedback $D_i$ and factors $I_i$ in $r_i$. & Eq.~\eqref{eq:reliance_new} \\
$r_i^*$ & History-smoothed reliance & Momentum-smoothed reliance. & Eq.~\eqref{eq:reliance_update} \\
$\alpha$ & Momentum weight & Balances historical $r_i^*$ and new $r_{i+1}$ in the update. & Eq.~\eqref{eq:reliance_update} \\
$\hat{r}$ & Reliance threshold & Evaluate whether human trust AI model. & AI reliance assessment \\

\midrule
\multicolumn{4}{@{}l}{\textit{AI Performance Assessment}}\\
$D_i$ & AI Performance feedback & Human evaluation of model predictions on $T_i$. & Eq.~\eqref{performance_assessment_model}, ~\eqref{performance_assessment_human} \\
$d_i^m$ & Model-only evaluation score & Evaluation when only model output is available. & Eq.~\eqref{performance_assessment_model} \\
$d_i^h$ & Human–model evaluation score & Evaluation when both human and model outputs are available. & Eq.~\eqref{performance_assessment_human} \\
$c$ & Scaling coefficient & Scales $d_i^m$ or $d_i^h$ to form $D_i$. & Eq.~\eqref{performance_assessment_model}, ~\eqref{performance_assessment_human} \\
$\theta^m,\theta^h$ & Satisfaction thresholds & Acceptance thresholds for AI performance assessments. & AI performance assessment \\
\midrule
\multicolumn{4}{@{}l}{\textit{Model-irrelevant factors}}\\
$I_i$ & Model-irrelevant factor & Sum of model-irrelevant influences on reliance. & Eq.~\eqref{model-irrelavant} \\
$C_i$ & Self-confidence term & Effect of self-confidence on reliance. & Eq.~\eqref{eq:self-conf} \\
$K_i$ & Risk term & Effect of task risk on reliance. & Eq.~\eqref{eq:risk} \\
$O_i$ & Complexity term & Effect of task complexity on reliance. & Eq.~\eqref{eq:complexity} \\
$S_i$ & Time-sensitivity term & Effect of time-sensitivity on reliance. & Eq.~\eqref{eq:sensitivity} \\
\bottomrule
\end{tabularx}
\label{tab:vars-timing}
\end{table*}

\begin{figure}[!tb]
\centering
\includegraphics[width=0.5\linewidth]{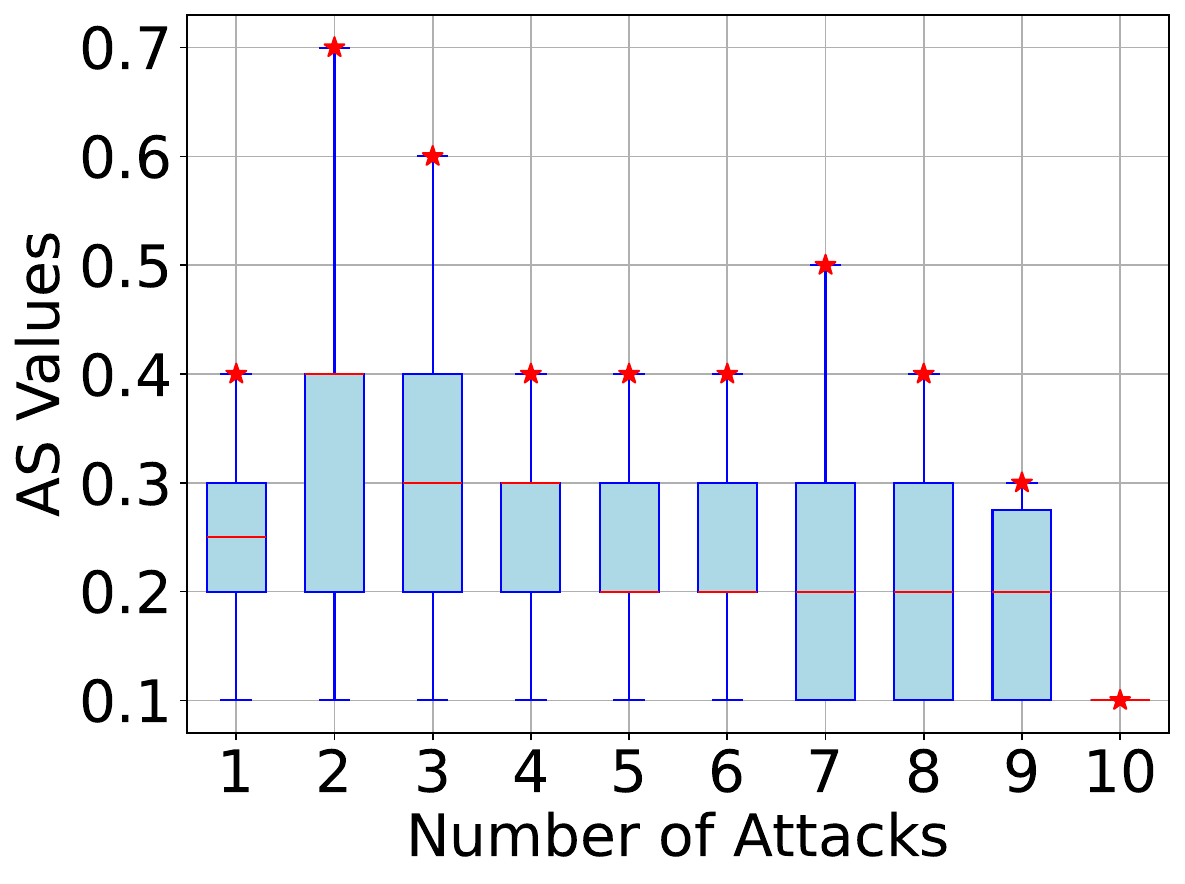}
\caption{
        The distribution of Attack Scores (AS) for varying numbers of attacks in 10 tasks.
    }
    \label{fig:attack_score_dist}
    \vspace{-1em}

\end{figure}

\section{Simulation Analysis}
\label{sec:simulation}

We conduct a comprehensive simulation study of human-AI decision-making tasks to assess how attack strategies behave under different settings. In this simulation analysis, we assume the AI model's accuracy is $p_m = 0.8$, which means the rate of error of model $e^M$ is 20\%; the human's accuracy is $p_h = 0.9$, which means the rate of error of human $e^H$ is 10\%\footnote{We set human accuracy slightly above model accuracy (\(p_h>p_m\)) as a conservative, safety-oriented assumption. The reason is that in most tasks that can replace humans (e.g., autonomous driving, object classification, etc.), humans are more likely to make correct judgments than AI models. This assumption is not universal; we also report sensitivity analyses that vary \(p_h\) and \(p_m\) in~\Cref{sensitivity_analysis}.}. On each task, the model and the human form their judgments \emph{independently}; their errors are independent draws with rates \(e^M\) and \(e^H\), respectively. The reliance update weight $\alpha = 0.8$, the scaling coefficient to calculate the new reliance $c = 1$. The threshold for AI reliance assessment $\hat{r} = 0.7$\footnote{We also provide a sensitivity analysis over \(\hat{r}\) in~\Cref{sensitivity_analysis}, reflecting variability in human reliance thresholds.} and initial reliance $r_1 = 0.8$ so that the human will automatically trust the first task (i.e., \(r_1 > \hat{r}\)). For updating the reliance value in AI performance assessment, we set evaluation score $d_i$ based on trust: if the AI is \emph{not} trusted and human override, \(d_i \sim \mathrm{Unif}[0,0.3]\); otherwise, \(d_i \sim \mathrm{Unif}[0.7,1]\).
We assume that the adversary can conduct arbitrary numbers of attacks ranging from $0$ to $10$ times.

\subsection{Impact of the Number of Attacks}

We investigate how the number of adversarial attacks affects overall attack success.~\Cref{fig:attack_score_dist} shows the distribution of Attack Scores (AS) across different attack frequencies over 10 tasks. Each box represents the spread of AS values, while the red markers indicate the best-performing strategy for each number of attacks, which is the highest AS value achieved at that point.

\textbf{Surprisingly, increasing the number of attacks does not always lead to higher attack scores (AS).} In fact, the most effective attack strategy occurs when only two attacks are deployed in this setting. After this point, both the median and maximum AS values tend to plateau or even decline. This non-monotonic trend highlights a key behavioral insight: repeated exposure to incorrect or inconsistent AI outputs can undermine user trust. As the number of adversarial interventions increases, humans may become more skeptical of the AI system and shift toward independent decision-making. This shift reduces the effectiveness of subsequent attacks, as the human begins overriding the AI rather than deferring to it.

Moreover, the relatively stable distribution of AS values beyond 2–3 attacks suggests diminishing returns when attacks are too frequent. Attacks deployed too frequently may even become counterproductive, as humans increase the chance of detection or raise suspicion. Thus, adversaries may benefit most from strategic, sparse attacks that exploit early-stage trust before user skepticism is triggered.

\subsection{Analysis of Attack Strategies in Two-time Attacks}

Given the observed effectiveness of two-time attacks, we analyze the best and worst attack strategies. To ensure that reliance values and AS values evolve consistently across each task, we fix the evaluation score \(d_i=0.3\)  for tasks subjected to attacks and \(d_i=0.7\) for non-attacked tasks, while holding the human prediction \(y^h\) and model prediction \(y^m\) constant.

\begin{figure}[!tb]
    \centering
    \subfloat[Reliance Value $r_i^*$\label{fig:reliance}]{
        \includegraphics[width=0.6\linewidth]{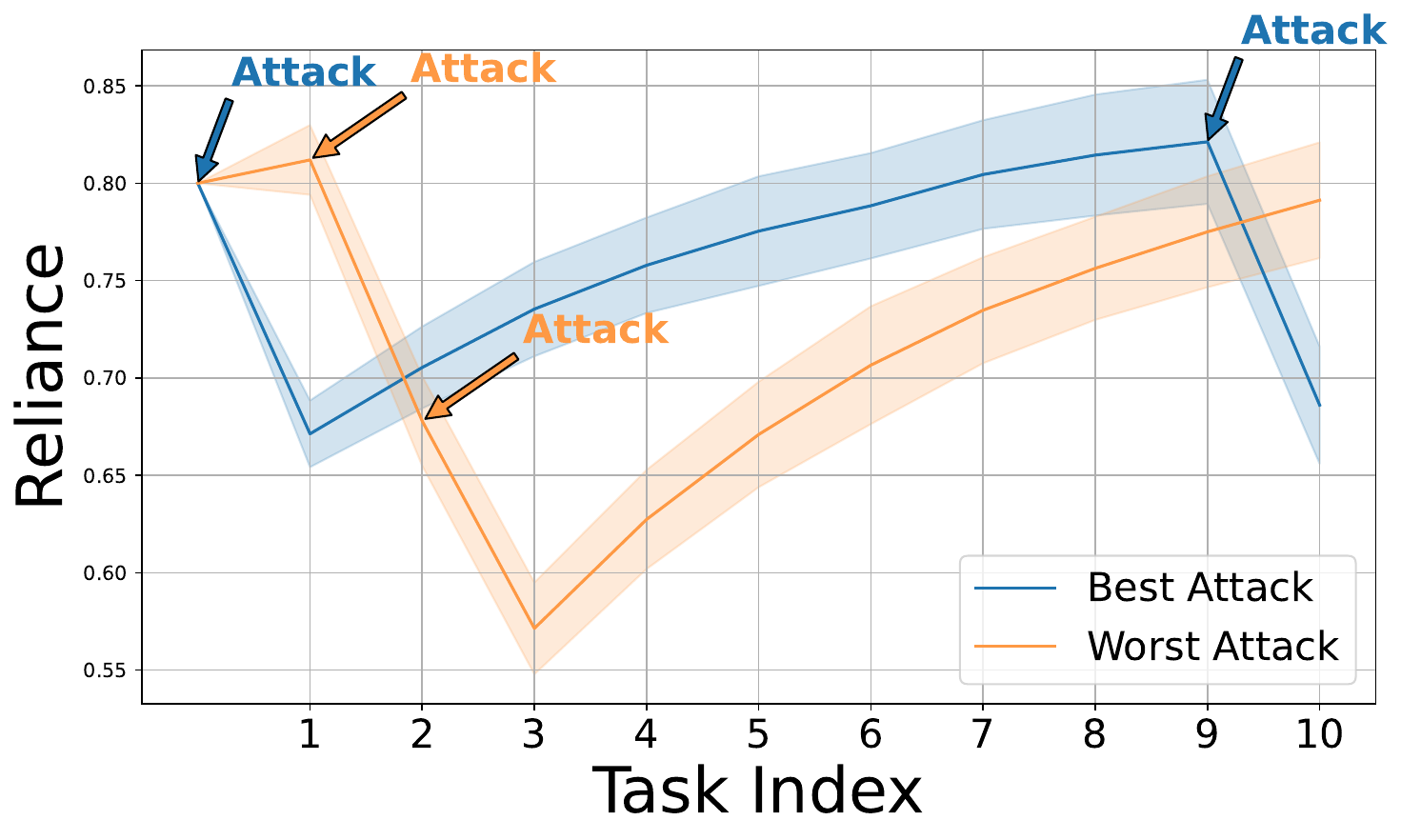}
    }

    \subfloat[Performance Assessment Score\label{fig:evaluation_score}]{
        \includegraphics[width=0.6\linewidth]{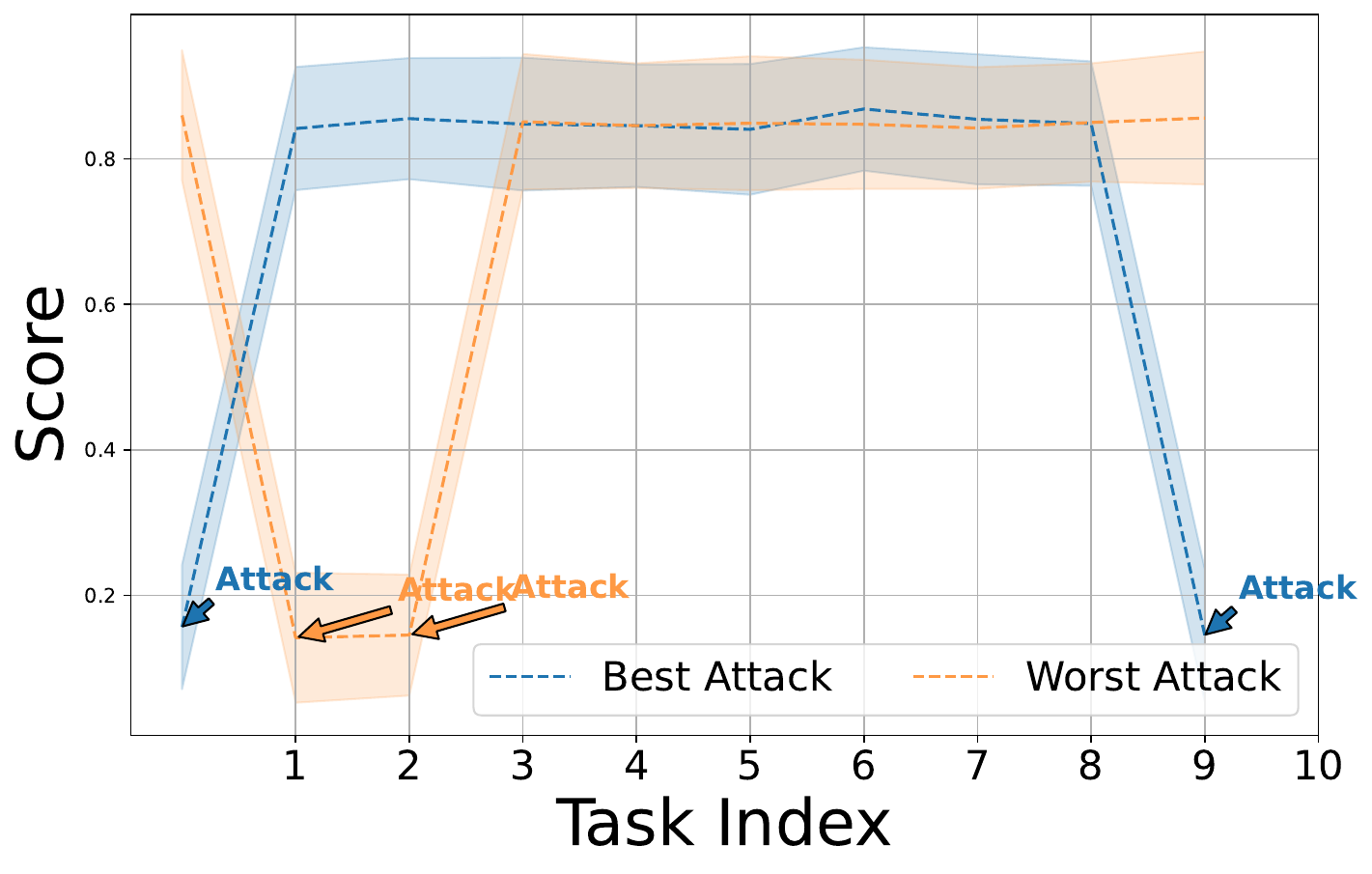}
    }

    \caption{
        (a) Reliance values and (b) performance assessment scores under the most (blue) and least (orange) effective attack strategies across 10 tasks. Shaded regions represent variance across runs, and annotated arrows show the timing of injected attacks. 
    }
    \label{fig:reliance_evaluation}
    \vspace{-1em}

\end{figure}

\Cref{fig:reliance} and \Cref{fig:evaluation_score} show that distributing attack timing separately yields better results than placing them consecutively. After the first attack, reliance is partially restored in one of the middle tasks, increasing the effectiveness of the second attack. \textbf{In contrast, the worst-case scenario of continuous attacks drives reliance to its lowest point and leaves the human consistently overriding the AI model. The result aligns with our previous analysis in~\Cref{case study}.} The optimal scenario is to carry out attacks at the beginning and end of the tasks. By providing accurate results in the middle tasks, the AI model helps humans regain reliance, ultimately enabling the success of the final attack. Conversely, the worst-case scenario occurs when two consecutive attacks cause humans to completely lose confidence in the model.

\subsection{Sensitivity Analysis}
\label{sensitivity_analysis}

\begin{figure}[!tb]
    \centering

    \subfloat[Impact of AI Model Accuracy ($p_m$).\label{fig:subfig1}]{
        \includegraphics[width=0.7\linewidth]{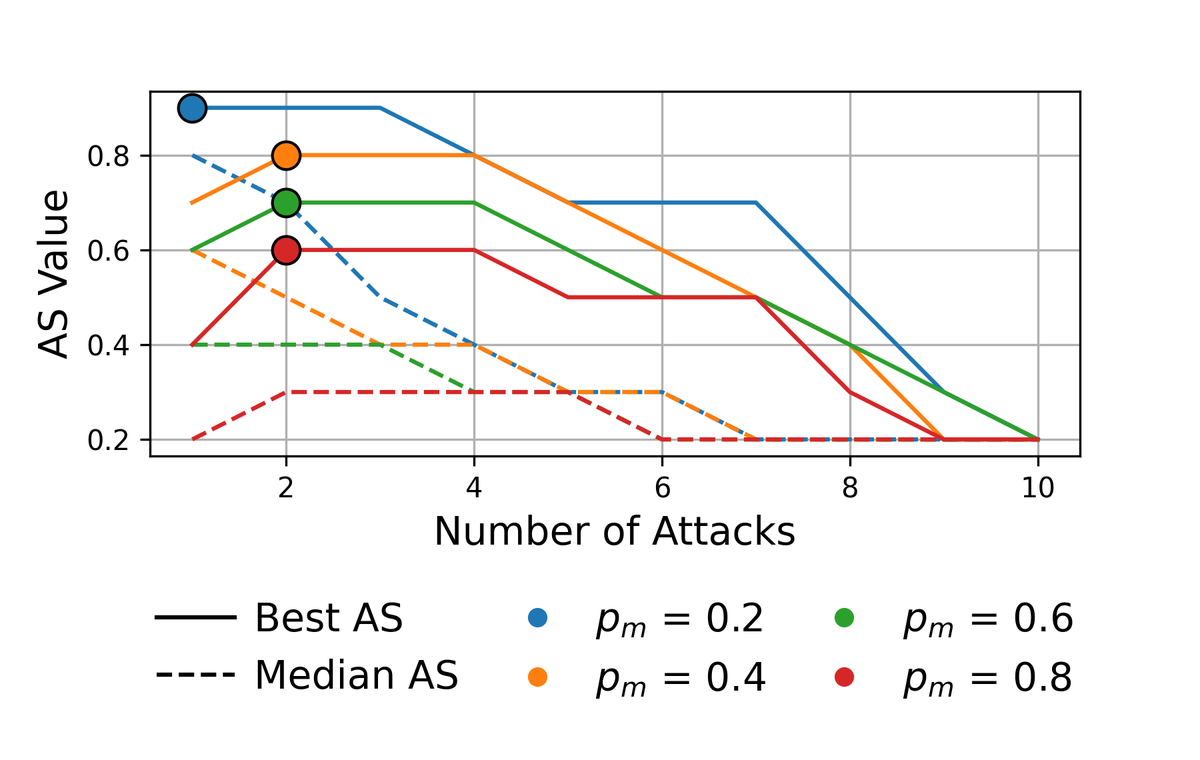}
    }

    \subfloat[Impact of Human Accuracy ($p_h$).\label{fig:subfig2}]{
        \includegraphics[width=0.7\linewidth]{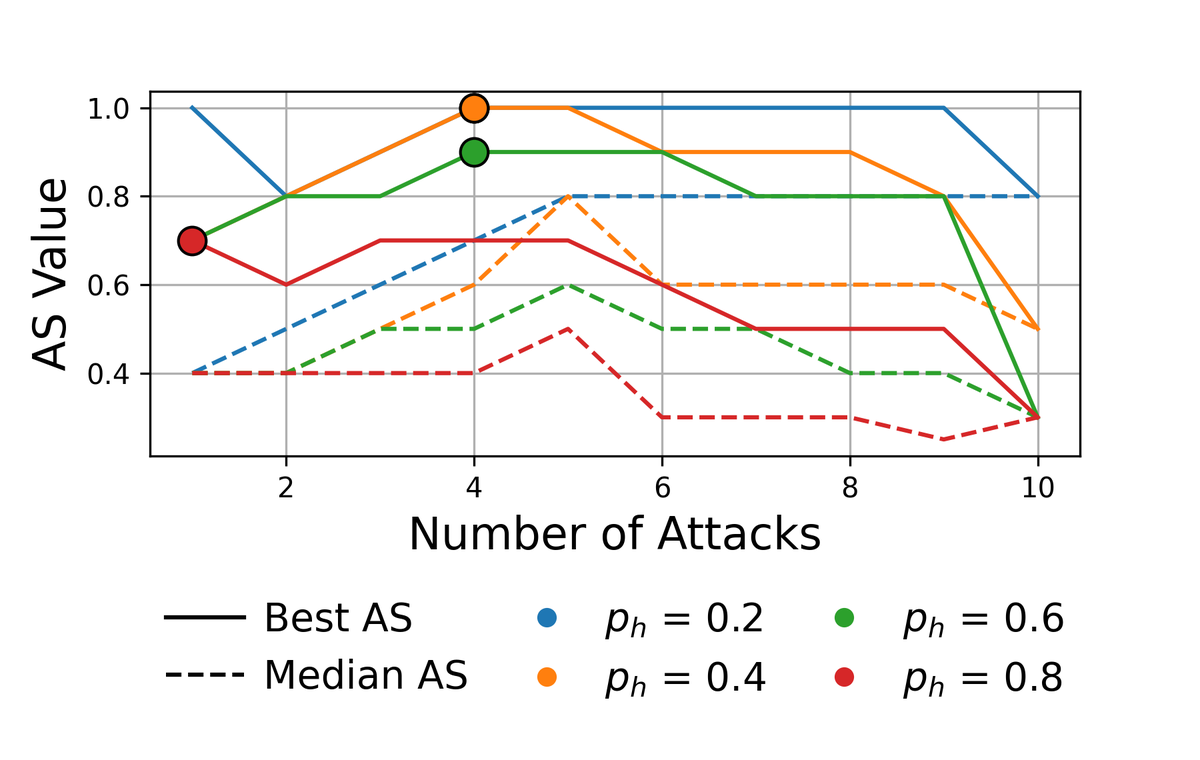}
    }

    \subfloat[Impact of Combined Model and Human Accuracy.\label{fig:subfig3}]{
        \includegraphics[width=0.7\linewidth]{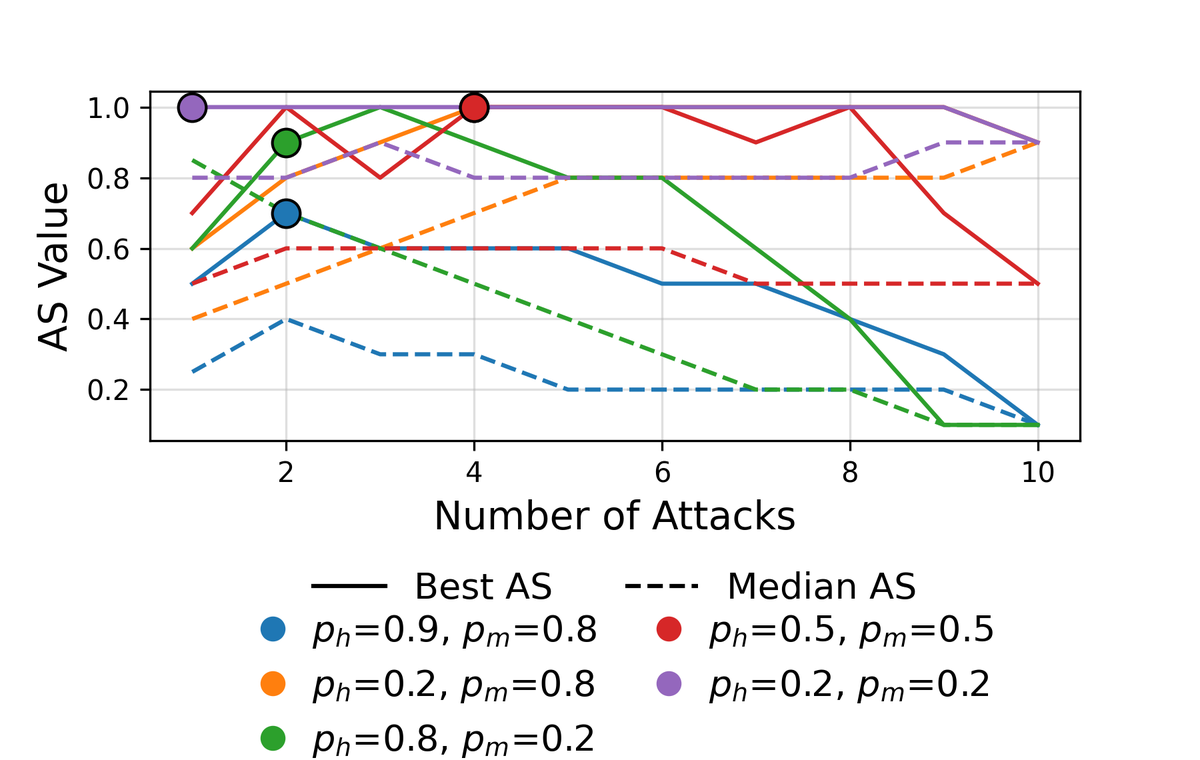}
    }

    \subfloat[Impact of Reliance Threshold ($\hat{r}$).\label{fig:subfig4}]{
        \includegraphics[width=0.7\linewidth]{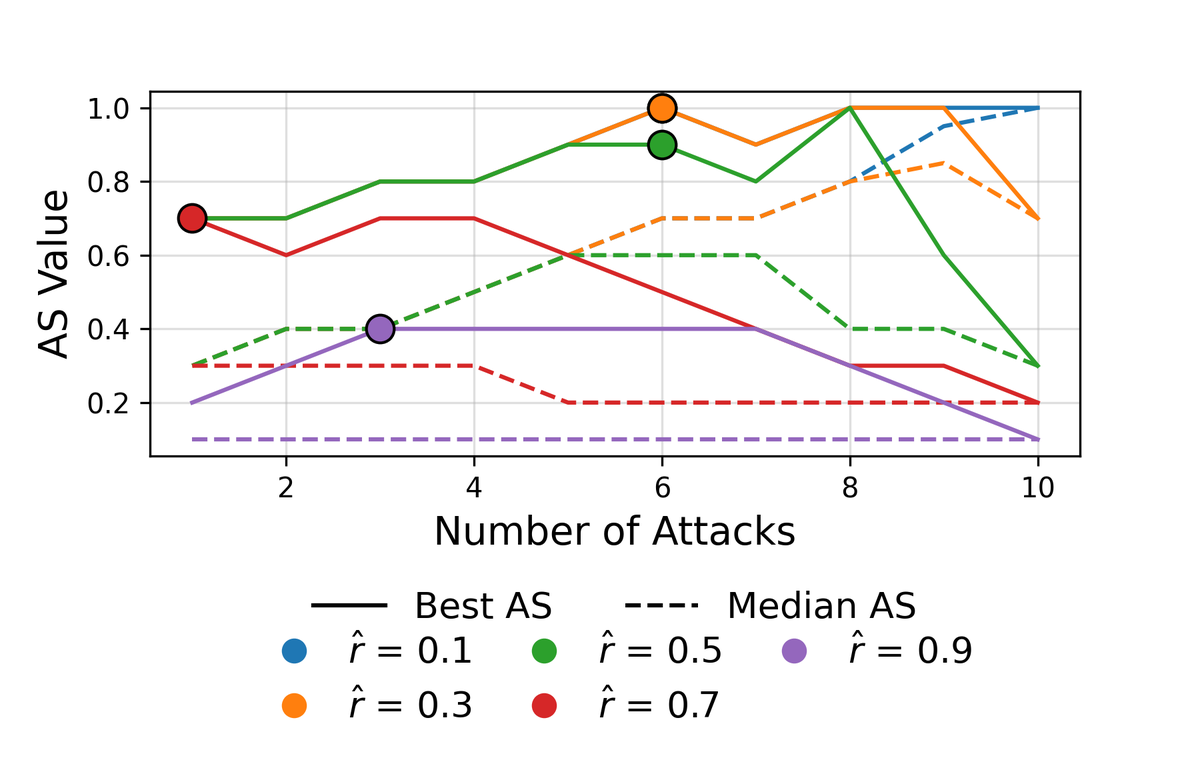}
    }

    \caption{Sensitivity analyses on AI model accuracy ($p_m$), human accuracy ($p_h$), combined model and human accuracy, and the reliance threshold ($\hat{r}$). Each subplot illustrates how the attack success (AS) value changes with the number of attacks under varying parameter settings. Highlighted dots indicate the optimal number of attacks that achieve the highest AS value for each condition.}
    \label{fig:4in1}
    \vspace{-1em}

\end{figure}

We conduct systematic sensitivity analyses to assess the robustness of our results in~\Cref{fig:4in1}. These analyses vary key parameters in the human-AI decision-making system, including AI model accuracy (\(p_m\)), human accuracy (\(p_h\)), and reliance threshold (\(\hat{r}\)). Model accuracy and human accuracy refer to the prediction accuracy of the AI model and the human, respectively; the reliance threshold determines when humans choose to accept AI outputs without seeking additional validation. Each plot shows how the attack success (AS) value changes as the number of attacks increases under different parameter settings. The key findings include:

\begin{itemize}
  \item \textbf{Model Accuracy (\(p_m\))}: \textit{Higher AI accuracy lowers attack success rates but does not significantly affect the system’s sensitivity to the number of attacks.} In practice, it is undeniable that AI models exhibit varying levels of predictive accuracy. For example, Med-PaLM~\cite{singhal2023large} demonstrates strong performance in medical diagnosis, achieving expert-level accuracy on benchmark clinical datasets, whereas general-purpose models like GPT-3.5 struggle with domain-specific medical reasoning~\cite{nori2023capabilitiesgpt4medicalchallenge}. By varying model accuracy \(p_m\) from 0.2 (representing weaker models) to 0.8 (representing stronger models) with fixed human accuracy \(p_h\) in~\Cref{fig:subfig1}, we observed that higher model accuracy leads to a monotonic decrease in attack success rate~(AS values), indicating that more accurate AI predictions reduce the adversary’s ability to succeed. However, even strong models (\(p_m = 0.8\)) remain vulnerable under multiple coordinated attacks. Notably, the optimal number of attacks (highlighted) varies slightly but shows no abrupt shifts across accuracy levels. 
  
  \item \textbf{Human Accuracy (\(p_h\))}: \textit{Lower human accuracy amplifies vulnerability, underscoring the importance of human oversight.} Obviously, human expertise also plays an essential role in the system’s resilience~\cite{nourani2020role,topol2019high,romeo2025exploring}. For instance, domain experts in medical diagnosis or legal reasoning are more likely to catch AI errors, whereas lay users may defer to incorrect model outputs~\cite{kostick2022ai}. With fixed model accuracy \(p_m\), varying human accuracy \(p_h\) from 0.2 (human without expertise) to 0.8 (human with expertise) in~\Cref{fig:subfig2}, demonstrating that weaker human judgment greatly increases system susceptibility. This highlights that adversaries can exploit human misjudgment, and that reliable human oversight is essential for mitigating risk.
  
  \item \textbf{Combined Accuracy}: \textit{The system is most vulnerable when both model and human accuracy are low, with human accuracy exerting stronger influence.}~\Cref{fig:subfig3} explores combinations of model accuracy \(p_m\) and human accuracy \(p_h\). The most severe vulnerabilities occur when both are low (i.e., \(p_m = 0.2\), \(p_h = 0.2\), representing the human without expertise using weaker models), whereas high human accuracy (\(p_h = 0.8\)) can compensate for imperfect AI performance. The trends suggest a nonlinear interaction where accurate human oversight can interrupt adversarial attack chains even when AI predictions are unreliable.
  
  \item \textbf{Reliance Threshold (\(\hat{r}\))}: \textit{Lower reliance thresholds (i.e., higher default trust in AI outputs) increase vulnerability to attacks, showing this parameter is critical to robustness.} Individuals exhibit varying levels of cognitive sensitivity when deciding whether to rely on AI outputs~\cite{buccinca2021}. For instance, some drivers using an AI-assisted navigation system may unquestioningly follow all route suggestions, even in unfamiliar or risky conditions, whereas others may double-check unexpected detours before accepting them. Thus, we analyze reliance thresholds \(\hat{r}\) ranging from 0.1 (users easily defer to AI) to 0.9 (users require stronger confidence to rely on AI), as shown in~\Cref{fig:subfig4}. The results show that a low reliance threshold, i.e., humans tending to trust AI, cause users to over-trust AI, magnifying adversarial success. Conversely, higher thresholds enforce human verification more often, improving overall system robustness. This indicates the reliance threshold \(\hat{r}\) is a critical parameter for the robustness of human-AI decision-making systems.
\end{itemize}

Overall, these sensitivity analyses highlight that adversarial vulnerability in human-AI decision-making systems is shaped not only by AI performance but also emerges from the interplay between model accuracy, human oversight, and behavioral dynamics such as reliance thresholds. While improving model accuracy reduces baseline susceptibility, human accuracy and cautious reliance behavior have a disproportionately large impact on adversarial robustness. Moreover, the presence of an optimal “attack window” suggests that timing and trust dynamics can be strategically exploited by adversaries but also potentially disrupted by adaptive defenses. These findings emphasize the need for holistic defense strategies that integrate system-level design, human behavior modeling, and dynamic trust calibration mechanisms.

\section{Potential Defense Strategies for Human-AI Collaboration}
\label{defense}
While this work primarily focuses on modeling and analyzing adversarial attack strategies in human-AI decision-making systems, we outline several potential defense directions for future exploration. 

\subsection{Human-in-the-Loop Defenses}
Unlike fully autonomous systems, the human-AI decision-making systems offer the unique advantages of active human participation in defensive strategies. Rather than relying solely on algorithmic safeguards, human-AI collaborative systems can incorporate human judgment, contextual awareness, and adaptive reasoning to detect and mitigate adversarial threats. However, this human-in-the-loop potential remains underutilized in existing defense paradigms. One promising direction involves leveraging user-initiated verification mechanisms. For example, users can request AI-generated explanations (e.g., using large language models as explainers~\cite{zhao2024,krause2024datacommonsensereasoninguse,bilal2025llms,cambria2024xai}) to verify whether AI predictions align with plausible reasoning. Such interactive scrutiny enables humans to catch inconsistencies or subtle anomalies that conventional automated methods may overlook. Prior research has shown that users interpret and rely on explanations differently depending on their expertise, confidence, and task familiarity~\cite{buccinca2021,ford2022explaining}. The personalized explanations may help align with user trust calibration~\cite{naiseh2023different,shin2021effects}. These hybrid systems may effectively reduce susceptibility to subtle adversarial manipulations. Importantly, future work could explore how to formalize these human-in-the-loop interactions into adversarial training pipelines or incorporate them into threat modeling frameworks.

\subsection{Defending Against Behavior-Exploiting Attacks}
As demonstrated in the paper, adversaries can exploit predictable human behavioral patterns (such as consistent reliance thresholds or self-confidence levels in model-irrelevant factors) to launch more severe and effective attacks. To mitigate such threats, a key defense direction involves preventing the exploitation of human behavior. This includes detecting anomalous interaction patterns that may indicate adversarial probing, obfuscating and randomizing consistent behavioral signals to hinder behavior modeling, and employing privacy-preserving techniques to ensure that human behavior remains difficult to infer or leak. For instance, an autonomous vehicle system could introduce randomized cross-checks between map data and visual perception to obscure consistent driver reliance patterns, making it harder for adversaries to predict when manipulated signs will be trusted without human verification.

\subsection{Theoretical Approaches to Robustness in Human-AI Systems}
Traditional theoretical robustness methods (e.g., certified robustness via randomized smoothing~\cite{cohen2019certifiedadversarialrobustnessrandomized,lyu2024adaptiverandomizedsmoothingcertified}) often assume deterministic models and fully autonomous systems, making them ill-suited for settings that involve human unpredictability. In contrast, human-AI systems need a broader modeling framework that accounts for the dynamic, non-stationary and context-sensitive nature of human behavior that challenges traditional robustness guarantees. In this context, game-theoretic approaches, such as strategic machine learning~\cite{ehrenberg2024adversariesincentivesstrategicalternative,rosenfeld_2024}, can be promising alternatives. The frameworks explicitly model strategic and dynamic interactions between adversaries, human users, and AI agents, offering a principled pathway toward robustness guarantees under behavioral uncertainty.

\subsection{Closing the Loop: Toward Integrated Defense Pipelines}

Ultimately, robust human-AI systems will require a coordinated integration of multiple defense layers: spanning user interface design, behavior-aware detection, model-level robustness, and theoretically grounded guarantees. An important future direction lies in designing unified pipelines that continuously monitor, adapt, and defend across these levels. Such pipelines can dynamically adjust trust calibration, modify explanation exposure, or throttle risky decisions based on real-time assessments of adversarial likelihood and human susceptibility. Moving toward such comprehensive, context-aware defenses will be critical for ensuring the resilience of next-generation AI systems deployed in high-stakes, collaborative settings.

\section{Discussion and Challenges}
\label{sec:discussion}

This work presents a human behavior-aware perspective on adversarial attacks in human-AI collaborative systems, emphasizing how dynamic human reliance creates novel vulnerabilities. While our findings shed light on key mechanisms and strategy insights, several challenges remain open for future investigation.

\subsection{Static Behavior Modeling}

Current approaches to modeling human reliance rely on simplified assumptions: e.g., four model-irrelevant factors, fixed reliance update rules, unrealistic linear modeling method, etc. However, human reliance in real-world settings is far more heterogeneous and context-sensitive~\cite{kaufman2025,lee2004trust}. Individual traits (e.g., confidence, cognitive load tolerance), domain expertise, and prior experience can significantly shape how reliance evolves~\cite{nourani2020role,bauer2023expl}. However, capturing or estimating such personal characteristics remains difficult in practice. Additionally, different task types (e.g., diagnostic reasoning vs. translation) may evoke distinct reliance patterns. A major open challenge lies in developing fine-grained, adaptive models of human behaviors that can support more realistic simulations and adversarial analysis.

\subsection{Balancing Robustness and User Experience}

Designing robust human-AI systems inevitably raises trade-offs between security and usability~\cite{braz2007designing,yang2020closer}. Defense strategies such as withholding uncertain output, issuing warnings, or requiring human verification can mitigate adversarial risks, but may also introduce friction, reduce efficiency, or harm user satisfaction~\cite{simkute2025ironies,honeycutt2020soliciting}. Overly defensive interfaces could even trigger unnecessary skepticism, undermining the benefits of AI assistance~\cite{dzindolet2003role,sunshine2009}. Future work must explore how to strike an optimal balance between adaptive robustness and smooth user experience, possibly through context-aware interventions that activate only when behavioral signals suggest that manipulation is likely.

\subsection{Multi-Agent Collaboration and Group Reliance}

Much of the current literature, including our work, focuses on single-user AI collaboration. Yet in practice, many decision-making settings involve multiple human stakeholders, e.g., clinical teams or financial analyst teams~\cite{hoch2023multi}. In multi-agent settings, reliance may exhibit emergent effects: reliance may be socially reinforced or diluted; adversarial misinformation may propagate through shared interfaces or majority influence~\cite{hammond2025multi,pastor2024large}. These collective dynamics introduce new vulnerabilities that are not captured by human-centric models. Future research should extend adversarial analysis to group-based collaboration, where adversaries may exploit consensus mechanisms, social proof, or communication asymmetries.

\section{Conclusion}

In this paper, we advocate that human factors fundamentally reshape adversarial attacks in human-AI decision-making systems. We revisit the human-AI decision-making systems under adversarial robustness, identify the key human behavioral factors and introduce an innovative robustness analysis framework with two key components: \textit{AI reliance assessment} and \textit{AI performance assessment}. Within this framework, we analyze various adversarial strategies and compare them with traditional AI-only decision making. As a case study, we introduce a timing-based adversarial attack and conduct simulation experiments to evaluate its impact. Our results reveal that increasing attack frequency does not always enhance adversarial impact and that, given a limited number of attacks, strategically distributed attacks are more effective than consecutive ones. In addition, we further conduct a comprehensive sensitivity analysis to examine how specific parameters affect adversarial effectiveness. The findings highlight that adversarial impact is highly sensitive to the model’s prediction accuracy and the human's confidence calibration. 
Looking ahead, we outline several potential defense strategies to enhance human-AI collaborative system robustness and discuss several key open challenges in developing behavior-aware, scalable, and adaptive systems against adversarial threats under real-world constraints. As AI continues to be deployed in high-stakes, interactive settings, understanding and safeguarding the human component will be central to the next generation of adversarial defense.

\bibliographystyle{IEEEtran}

\bibliography{refs}    

\newpage
\appendices

\section{Literature Review for Human-Related Adversarial Analysis}
\label{app:human-related_literature}

To support our position and provide concrete evidence, we conduct a systematic analysis of the existing adversarial attack and defense literature (2021–2025) to assess the extent of research addressing human-related concerns. We begin with the curated paper list maintained by Carlini (link: \url{https://nicholas.carlini.com/writing/2019/all-adversarial-example-papers.html}), which aggregates available publications related to adversarial machine learning on arXiv across years, containing 9,275 papers. To identify papers potentially related to human factors, we first perform a keyword-based filtering process. 
To construct this filter, we select a set of 15 human-centric keywords including "human," “trust,” “perception,” "cognition," “behavior,” “decision-making,” “interaction,” “explanation,” "emotion," “social,” “psychology,” "mental," “confidence,” “usability”. These keywords were chosen based on their prevalence in the human-centered AI, human-computer interaction (HCI), and explainable AI (XAI) literature~\cite{schneider2018empowerment,capel2023human,rong2023towards}, and were designed to capture a broad range of factors that shape how humans interpret, trust, or interact with AI systems. This filtering step yielded a subset of 5,719 papers whose titles or abstracts included at least one of the selected terms.

After an initial filtering of adversarial machine learning papers related to human factors, we concentrated our analysis on six widely studied human-related categories: (1) human perception of perturbations, (2) trust in AI systems, (3) impact on decision making, (4) usability and user experience, (5) human-in-the-loop analysis, and (6) ethical and social implications, which are relevant to human-AI decision-making. These categories were chosen because they represent the primary ways in which adversarial attacks can affect, exploit, or be shaped by human cognition, behavior, and societal context~\cite{baniecki2024adversarial,noppel2024sok,sperrle2021survey,oprea2023adversarial}. To categorize papers into these themes, we computed the semantic cosine similarity between each abstract and predefined textual descriptions for each category using sentence embeddings~\cite{reimers2019sentence}. A similarity threshold of 0.4 was chosen to capture papers with moderate or stronger alignment to human-centered concerns, balancing coverage and precision. As shown in \Cref{table:human}, \textbf{only 9 of 9,275 papers (approximately 0.09\%) engage meaningfully with human-related topics} upon manual verification, underscoring how underexplored human factors remain in adversarial machine learning, despite their growing importance in human-AI systems.

\end{document}